%Paper: astro-ph/9406070
%From: frank@solaris.astro.uu.se (Frank Pijpers)
%Date: Mon, 27 Jun 94 21:00:37 +0200

%%%%%%%%%%%%%%%%%%%%%%%%%%%%%%%%%%%%%%%%%%%%%%%%%%%%%%%%%%%%%%%%%%%%%
%
%  AGN-SOLA article    MNRAS VERSION 1, 30 May 94, FPP
% !!!!!!!!!!!!!  uses  the MNRAS macro-package
%
%%==================================================================%%
%
%FPP some macros  :
%    automatic equation numbering :
\newcount\eqnumber
\eqnumber=1
\newcount\fignumber
\fignumber=1
\def\neqn{{\rm(\the\eqnumber)}\global\advance\eqnumber by 1}
\def\nfig{\global\advance\fignumber by 1}
%    referring to equations and figures by name :
\def\eqnam#1#2{\immediate\write1{\xdef#2{(\the\eqnumber}}
\xdef#1{(\the\eqnumber}}
\def\fignam#1#2{\immediate\write1{\xdef\ #2{\the\fignumber}}
\xdef#1{\the\fignumber}}
%  putting hats on symbols
\def\spose#1{\hbox to 0pt{#1\hss}}
\def\tophat{\spose{\raise 0.5ex\hbox{\hskip4.0pt$\widehat{}$}}}
\input mn.tex
\hyphenation{Pij-pers}
%FPP
%
%

%  To generate a "referee version" (single column, double-line spacing)
%  activate the command by deleting the "%" in the following line:
%\Referee
%\Autonumber
\loadboldmathnames

\begintopmatter

\title{Reverberation mapping of active galactic nuclei~: \hfil\break
The SOLA method for time-series inversion }
\author{ Frank~P.~Pijpers$^1$ and Ignaz~Wanders$^{1,2}$ }
\affiliation{$^1$ Uppsala Astronomical Observatory,
             Box 515, S-751\thinspace{}20 Uppsala, Sweden }
\affiliation{$^2$ Present address: Department of Astronomy, the Ohio State
             University, 174 West 18$^{\rm th}$ Avenue, Columbus, OH 43210,
USA}
\shortauthor{F.P. Pijpers and I. Wanders}
\shorttitle{SOLA time series inversion}
\acceptedline{Accepted for publication in MNRAS}
\abstract{
We present a new method to find the transfer function (TF)
of the broad-line region in active galactic nuclei. The subtractive
optimally localized averages (SOLA) method is a modified version of
the Backus-Gilbert method and is presented as an alternative to the
more often used maximum-entropy method for the inversion of variability data.
The SOLA method has been developed for use in helioseismology. It has
been applied to the solar oscillation frequency splitting data currently
available to deduce the internal rotation rate of the sun. The original
SOLA method is reformulated in the present paper to cope with the slightly
different problem of inverting time series. We use simulations to test
the viability of the method and apply the SOLA method to the real data
of the Seyfert-1 galaxy NGC\thinspace{}5548. We find similar TFs for these data
as previous studies using the maximum-entropy method. We
confirm thereby previous results while simultaneously presenting an
alternative and independent inversion method. Moreover, we do not find
significant negative responses in the TF. The integral of the TF, an
important quantity measuring the total observed line-processing in the
broad-line region, is correctly reproduced by the SOLA method with
high accuracy. We investigate the effects of measurement errors and how
the resolution of the TF critically depends upon both the sampling rate
and the photometric accuracy of the data.
}
\maketitle
%-----------------------------------------------------------------------

\section{Reverberation mapping}

The broad-line region (BLR) in active galactic nuclei (AGNs) is too
small to be resolved spatially with even the largest planned telescopes.
Indirect methods must therefore be employed to obtain information about
its structure and dynamics. One way of doing so is via reverberation
or echo mapping, a technique possible for variable sources only.

The broad emission lines originating in the BLR are photoionized by a
small central continuum source (see Netzer 1991 for a detailed review
of the standard model). In many AGN, especially the low-luminosity ones,
this continuum source is observed to be variable in its flux. The broad
emission lines respond to the continuum variations,
albeit with a time delay of several days due to light-travel-time effects
through the BLR. The high velocities of the clouds comprising the BLR
redistribute line photons in wavelength, resulting in the broad emission-line
profiles. Therefore, the combination of flux and profile variations of
the emission lines in response to the ionizing-continuum variations can
be used to map the phase space of the BLR (Blandford \& McKee 1982; see
Peterson 1993 and references therein for an extensive review).

Mathematically, the concept of reverberation mapping reduces to the inversion
problem
\eqnam\RevMapV{RevMapV}
$$
L(v,t)=\int{\rm d}\tau\ \Psi(v,\tau)C(t-\tau),
\eqno\neqn
$$
where $L(v,t)$ is the observed emission-line light curve, $v$ the
projected velocity with $v=0$ the line centre, $C(t)$ the observed
continuum light curve, and $\Psi(v,\tau)$ the transfer function (TF)
of the BLR. The TF is the quantity that holds the information about
the geometry, kinematics and physics of the BLR and is thus of central
importance. To obtain higher $S/N$ in the time series, Eq. \RevMapV) is
usually integrated over projected
velocity $v$ which reduces it to the total-flux case
\eqnam\RevMapC{RevMapC}
$$
L(t)=\int{\rm d}\tau\ \Psi(\tau)C(t-\tau).
\eqno\neqn
$$
To date two inversion methods have been applied to AGN variability data to
invert Eq. \RevMapC) and so deduce the TF. These are the Fourier-transform
method (FTM) (e.g., Bracewell 1986; Maoz et al. 1991) and the
maximum-entropy method (MEM) (Skilling \& Bryan 1984;
Horne, Welsh \& Peterson 1991; Krolik et al. 1991). The FTM is very
noise-sensitive and cannot handle irregularly sampled light curves
properly. Only for the galaxy NGC\thinspace{}4151 has a TF been derived
via this method (Maoz et al. 1991) but without a clear result.
The MEM on the other hand has been successful in finding a number
of TFs for several AGN: NGC\thinspace{}5548 (Horne, Welsh \& Peterson 1991;
Krolik et al. 1991; Peterson 1993; Peterson et al. 1994), Mkn\thinspace{}590
(Peterson et al. 1993) and NGC\thinspace{}3516 (Wanders \& Horne 1994).
A third method, based on a linear regularization method (cf. Press
et al.~1992) is currently under development (Krolik 1994).

In this paper we will outline another method for the inversion of Eq.
\RevMapC), not hitherto applied to AGN. This method is a modified version
of the Backus-Gilbert method (Backus \& Gilbert 1967, 1968, 1970) in
use in geoseismology. More recently the method has been applied to
helioseismology, specifically to the solar-oscillation frequency-splitting
data to deduce the internal rotation rate of the sun. A comparison of
various methods in use in this field, including the Backus-Gilbert
method, can be found in papers by Gough (1985) and by Christensen-Dalsgaard,
Schou \& Thompson (1990). The Backus-Gilbert method is also sometimes
referred to as optimally localized averages (OLA) or multiplicative OLA
(MOLA). The MOLA methods are generally considered not to be cost-effective
since they require numerous matrix inversions of large matrices. This
led Pijpers \& Thompson (1992, 1994) to reformulate the OLA method into
a fast method called the subtractive optimally localized averages (SOLA)
method. In Sect. 2 this method will be outlined in more detail and the
transformation from the helioseismological formulation to the
reverberation-mapping concept will be made. Section 3 describes on which
points the application to reverberation mapping differs from helioseismology,
and problems that are specific to AGN time-series analysis will be
addressed. Section 4 discusses simulations done on realistic (i.e.
noisy and irregularly sampled) artificial data and shows that recovery
of the TF by the SOLA method can be done with currently available data.
In Sect. 5 we then apply the SOLA method to the real data of
the Seyfert-1 galaxy NGC\thinspace{}5548 and present its TF for four years
of monitoring data. We show that the presently derived TFs are
consistent with the previously derived ones using the MEM and the
observed global changes in the TF from year to year are reproduced
by the SOLA method. These global changes are both changes in the
position of the peak of the TF and in the integral of the TF. The
evolution of the BLR in NGC\thinspace{}5548 on the dynamical time scale is
thereby confirmed. In Sect.~6 we summarize our results.

\section{The SOLA method}
The strategy of the SOLA method is to find a set of linear coefficients
which, when combined with the data, produce the unknown convolved
function under the integral sign. In the application to reverberation
mapping these coefficients will be multiplied with the individual line
fluxes in the measured time series and then summed to obtain an estimate
of the TF at a certain lag $\tau_0$.

Consider again Eq. \RevMapC) but now with the line fluxes explicitly
written as a time series consisting of $N$ individual measurements
$L(t_i)$ with an associated measurement error $\delta L (t_i)$
\eqnam\LFTimSer{LFTimSer}
$$
L(t_i) - \delta L (t_i) = \int\limits_{0}^{\infty}{\rm d}\tau\
\Psi(\tau ) C(t_i-\tau)\hskip 0.5truecm \forall\ i=1,...,N.
\eqno\neqn
$$
The lower limit of the integration is set to $0$ which implies that the
line flux is assumed to respond to the continuum variations and not vice versa.
If the continuum flux $C$ would be known for all values of $t$ and would
be error-free
this problem would be equivalent to the problem of inverting helioseismology
data. In that case an {\it averaging kernel} ${\cal K}$ would be
constructed~:
$$
{\cal K} ( \tau_0, \tau)\ =\ \sum\limits_{i=1}^{N} c_i(\tau_0) C(t_i-\tau)
\eqno\neqn$$
which is strongly peaked around $\tau_0$ and which is normalized~:
\eqnam\constraint{constraint}
$$
\int\limits_{0}^{\infty} {\rm d}\tau\ {\cal K} (\tau_0, \tau) \equiv 1.
\eqno\neqn$$
An estimate $\tophat \Psi (\tau_0)$ of the TF at time lag $\tau = \tau_0$ is
then obtained by
combining the line fluxes with this same set of coefficients
\eqnam\OldInv{OldInv}
$$
\eqalign{
\tophat \Psi (\tau_0) \equiv \sum\limits_{i=1}^{N} c_i(\tau_0) L(t_i)
= \int\limits_{0}^{\infty} &{\rm d}\tau\ {\cal K} (\tau_0, \tau) \Psi(\tau)
\cr
&+ \sum\limits_{i=1}^{N} c_i(\tau_0) \delta L(t_i). \cr}
\eqno\neqn$$
The first term on the right of Eq. \OldInv) is a weighted average of $\Psi$
in which ${\cal K}$ is the weighting function. The other term expresses
the propagation of the errors in the line fluxes. In the SOLA method
the aim is to make the kernel ${\cal K}$ resemble some chosen target form
${\cal T}$ while at the same time moderating the effect of the errors in the
line fluxes. This is achieved by minimizing
\eqnam\OldMin{OldMin}
$$
\int\limits_{0}^{\infty}{\rm d}\tau\ \left[ {\cal K}(\tau_0, \tau) -
{\cal T}(\tau_0, \tau) \right]^2 + \mu \sum\limits_{ij} E_{ij} c_i c_j
\eqno\neqn$$
subject to the constraint \constraint). Here ${\bf E}$ is the error
variance-covariance matrix of the observed line fluxes. $\mu$ is a trade-off
parameter which may be chosen according to the relative desirability of
making the first and second terms in \OldMin) small. The ideal target
function for the averaging kernel would be a delta function but it is
clear that an infinite resolution cannot be attained in practice. A
convenient form of a target function with an adjustable width is
\eqnam\tarfun{tarfun}
$$
{\cal T}\ =\ {1\over f \Delta} \exp\left[ - \left({\tau - \tau_0 \over
\Delta}\right)^2 \right].
\eqno\neqn$$
Here $f$ is a normalization factor which is introduced to make the total
integral of ${\cal T}$ equal to unity. If $\Delta \ll 1$ the factor $f$
reduces to $\sqrt{\pi}$.

The minimization leads to the matrix equation (Pijpers \& Thompson 1994)
\eqnam\mateq{mateq}
$$
{\bf A}\cdot {\bf c} (\tau_0 )\ =\ {\bf v} (\tau_0 ).
\eqno\neqn$$
Here ${\bf c} (\tau_0 )$ is the vector of linear coefficients which is
unknown, with an extra $(N+1)$th element which is a Lagrange multiplier.
The elements of the symmetric matrix A are~:
\eqnam\crocor{crocor}
$$
A_{ij}\ =\cases{
 \int_{0}^{\infty} {\rm d}\tau\ C(t_i-\tau) C(t_j -\tau) + \mu E_{ij}
\ \ \hfill (i,j \le N) \cr
\hfill\int_{0}^{\infty} {\rm d}\tau\ C(t_i -\tau) \hfill (i \le N, j=N+1) \cr
\hfill\int_{0}^{\infty} {\rm d}\tau\ C(t_j -\tau) \hfill (j \le N, i=N+1) \cr
\hfill 0 \hfill (i=j=N+1) \cr}
\eqno\neqn$$
The elements of the vector ${\bf v}$ are given by
\eqnam\corvec{corvec}
$$
v_i(\tau_0)\ =\cases{
 \int\limits_{0}^{\infty}{\rm d}\tau\ C(t_i -\tau) {\cal T}(\tau_0, \tau)
\hskip 0.5truecm\hfill (i\le N) \cr
\hfill 1 \hfill (i=N+1)\cr}
\eqno\neqn$$
Note that the elements of the matrix ${\bf A}$ do not depend on the point
$\tau_0$ at which an estimate of the TF will be determined. The matrix
has to be inverted only once. After this the different vectors
${\bf v}(\tau_0)$ can be combined with the inverse matrix with relatively
little computational effort, to scan the TF over the range of $\tau_0$
of interest. It should also be noted that if a different error weighting
$\mu$ is chosen the matrix elements of $A$ can be calculated quickly but
a new matrix inversion is required.

In practice one finds that there is a trade-off between the error
weighting $\mu$ and the resolution width $\Delta$. Increasing $\mu$
produces a decreasing error in the TF propagated from the line flux
measurements but the averaging kernel ${\cal K}$ will depart more and
more from the target form ${\cal T}$ unless $\Delta$ is also increased.
This departure of ${\cal K}$ from the prescribed form introduces a
`systematic error' in the determination of the TF which is undesirable.
A more detailed discussion of this trade-off can be found in the paper of
Pijpers \& Thompson (1994).

\section{Aspects of time-series analysis with the SOLA method}

\subsection{Adapting SOLA}

It is clear that for the problem of time series analysis discussed here
the original formulation of SOLA cannot work since the continuum fluxes
are known only for discrete points in a finite time interval. An extra
problem is that this time series for the continuum flux also has
measurement errors. These problems will be dealt with in several steps
leading to a working algorithm for SOLA inversion of time series.

\subsection{The finite extent of the continuum time series}

\fignam{\exaser}{exaser}
\beginfigure{1}
\vskip 6.2cm
\caption{{\bf figure \exaser} Example of a time series with 10 measurements.
The index $i$
counts the consecutive measurements. The whole time series is re-plotted
as a function of delay time $\tau$ for each added measurement, with an
arbitrary vertical offset. The dashed lines represent the flux before the
first measurement in the series}
\endfigure\nfig
In Fig. \exaser{} an example is shown of a time series of measurements
of the continuum flux. The index $i$ counts the measurements. For each
added measurement the whole time series is replotted with an arbitrary
vertical offset. The time series is plotted as a function of delay time
$\tau$ so that the most recent measurement is always the left-most
point at $\tau = 0$. The dashed lines represent the unknown part of
the light curve before the first measurement. For this time series
of 10 measurements the 10 curves shown in Fig. \exaser{} would be
the set of base functions $C$ that appear in Eqs. \crocor) and
\corvec). From Fig. \exaser{} it is clear that it is not possible
to evaluate the integrals since there is always a part of the base
functions $C$ which is not known since it falls before the first
measurement. The solution to this problem is to decide before starting
the inversion to put the upper limit of the integrations to a certain
finite time lag $\tau_{\rm max}$. This implies that the TF will be assumed
to be negligible for time lags $\tau > \tau_{\rm max}$. In essence this
is an assumption regarding the maximum size of the BLR. If this parameter
is unknown one must experiment with various values for $\tau_{\rm max}$
and re-do the inversion for each value.

The maximum time lag $\tau_{\rm max}$ is indicated in Fig. \exaser{}
by the vertical dash-dotted line. Even with a finite value of $\tau_{\rm
max}$ there will still be some base functions that are known for only
part of the range $[0, \tau_{\rm max}]$ so these will have to be dropped
completely. This reduces the number of independent base functions. It
is important not to choose $\tau_{\rm max}$ too large because then the
number of independent base functions becomes very small which has an
adverse effect on the error in the reconstructed TF and also on the
resolution that can be attained.

It is also important not to choose $\tau_{\rm max}$ too small since
then each base function is determined by only a very few points. This means
that the integrals in \crocor) and \corvec), which are calculated on the
basis of an interpolation, become uncertain. Note that the discarding
of some of the base functions does not mean that the first continuum
measurements in the series are discarded. These are all used but shifted
to the right in the base functions for higher $i$. However, if it is
assumed that the measurements of continuum flux and line flux are always
simultaneous there is some data that is obtained observationally which
cannot be used and must be discarded. If the first $M$ base functions
are discarded the first $M$ line-flux measurements are also discarded.
Since the line flux of these first $M$ measurements (partly) responds to
variations in a part of the continuum flux that has not been measured,
these measurements are of limited use for the inversion. Because
the TF is only calculated at one $\tau_0$ at a time, one could,
in principle, use a floating $\tau_{\rm max}$, i.e., $\tau_{\rm max}$
is short if $\tau_0$ is short, such that fewer emission-line data are
ignored. However, this would mean the matrix inversion \mateq) must be
done for each different $\tau_{\rm max}$ which is expensive. Furthermore,
with finite resolution, boundary effects play a significant r\^ole when
$\tau_{\rm max}$ becomes of the order of twice the resolution or less.
All in all, it is not advantageous to do this and in this paper we will
keep $\tau_{\rm max}$ fixed and drop the first $M$ line-flux measurements.

\subsection{The discrete sampling of the continuum time series}

Even after limiting the inversion to a finite interval in $\tau$ the
base functions are not completely determined. The continuum flux is
only known on a discrete set of points. The second step towards a
working algorithm is to interpolate between the individual measurements.
It is important to realize that this has an impact on the resolution
that can be attained in the determination of the TF. It implies
that there is a minimum to the width $\Delta$ of the target function
\tarfun) even with error-free data.

At this point it is useful to discuss the `intrinsic' resolution that
could be attained with a series of error-free continuum-flux measurements.
If the sampling rate of the continuum flux is strictly regular the
base functions also are sampled strictly regularly. It is immediately
clear that the minimum possible resolution is then of the order of the
time interval between samplings. If the sampling is irregular the result
is less obvious. It is useful to distinguish at this point between
a regular time series with `missing data' and a fully irregular or
non-redundant time series.

In a time series of the first type the sampling can be thought of as
taken on integer multiples of some minimum time interval $\Delta t_{\rm
min}$. The `missing data' are then all those integer multiples of
$\Delta t$ for which a measurement is not available. In this case in
the diagram of Fig. \exaser{} a set of $K$ regularly spaced vertical
lines can be drawn with $K = \tau_{\rm max} / \Delta t_{\rm min}$. All
measured points of all the base functions will fall on those $K$ lines.
Considered in this way the minimum resolution width $\Delta$ with this
type of irregular sampling must be of the order of $\Delta t_{\rm min}$.
If a fraction $f$ of the data is missing in a time series with this
imaginary sampling rate $\Delta t_{\rm min}$ the length of the total
time series must be larger by a factor $1/f$ than a similar time series
without missing data, in order to really attain something close to this
resolution.

In a non-redundant time series there is no minimum sampling interval
for which all measurements fall on integer multiples of that time
interval. Note that this does not require that some samples are taken
with an infinitely small time interval between them. Consider again
Fig. \exaser{}, where now the interval between $0$ and $\tau_{\rm max}$
is rescaled so that it is mapped onto the interval $[0,1]$. In the limit
of an infinitely long non-redundant time series the set of $\tau /
\tau_{\rm max}$ for which at least one measurement is available would
lie dense in the unit interval. In other words the distribution of these
points would have the same properties as the rational numbers on the
unit interval. For an error free time series of this type the minimum
resolution width is inversely proportional to the total length of the
time series.

The time series of measurements of fluxes of variable AGN as they are
usually reported fall into the first category of irregularly sampled
data. The minimum time interval here is 1 day. However, one should not
expect to attain this limiting resolution of 1 day for two reasons.
The first reason is the finite error in the measurements of the line
flux. This leads to the usual trade off between errors in the TF and
resolution of the TF, mentioned in the previous section. The second
reason is finite error in the continuum flux measurements. These
errors give rise to an error or uncertainty in the integration kernel
${\cal K}$ that is constructed. This error increases for a decreasing
width of the target kernel ${\cal T}$ which effectively limits the
resolution width to a minimum which may lie well above the minimum
of 1 day for error-free data.

There is one other reason why this limiting resolution may not be
attained which has nothing to do with the sampling strategy or the
measurement errors. It is clear that if the continuum does not vary,
then it is impossible to deduce anything about a TF. As an extension
to this it is not hard to see that if there is a finite minimum time
scale on which the continuum source varies, which is larger than the
sampling time interval, one should not expect to attain a better
time resolution in the TF than that typical time scale of the continuum
source. This should in principle be reflected in an impossibility for the
algorithm to find a good kernel ${\cal K}$ with a smaller width than
this time scale (cf. Fig.~7).

\subsection{Interpolation of the continuum time series}

The practical problem left is to find an interpolation scheme for the
continuum flux time series so that the integrals of \crocor) and \corvec)
can be calculated. This interpolation must not introduce spurious
variations due to the errors in the continuum flux. When tested on
simulated time series a cubic spline was found to introduce spurious
oscillations with amplitudes of up to four times the formal errors on the
individual measurements. Such spurious structure in the continuum flux
would of course have no counterpart in the line-flux measurements.
In this way an artificial de-correlation would be found for some
determinations of the TF, effectively forcing the TF towards values
close to $0$. This would happen in particular for values of $\tau_0$
for which none or only a few base functions have a measured point.

The interpolation scheme decided upon here is a Savitzky-Golay smoothing
filter for the continuum time series, adapted to an irregular sampling
interval (cf. Press et al., 1992). With this algorithm a least-squares
fit of a low order polynomial to a moving window of $2N_{\rm s}+1$
points in the time series is constructed. Except at the beginning and
end of the entire series these points are always distributed symmetrically
around the point for which the polynomial will be used in the interpolation.
An interpolating polynomial constructed for the continuum light curve
around time $t_i$ is then used only in the interval $[(t_{i-1} + t_i)/2,
(t_i + t_{i+1})/2]$. For the $N_{\rm s}$ points at both ends of the time
series the number of points used is reduced one at a time until only
$N_{\rm s} + 1$ points are used to construct an interpolating polynomial for
the first and last point in the time series. In this case the order of
the polynomial is also reduced to make sure that the error in the
coefficients of the polynomial remains under control. A choice of a
moving window of 5 points (i.e. $N_{\rm s} = 2$) and a quadratic
polynomial are found to perform satisfactorily. At the boundaries the
polynomial was reduced to linear. The choice of the number of points in
the smoothing window and the degree of the polynomial in general must
depend on the ratio of the real point-to-point variations in the
continuum light curve and the statistical measurement noise.

Note that this interpolation scheme is not continuous for all $\tau$ in the
range $[0, \tau_{\rm max}]$, contrary to a cubic spline. This is of no
importance for the algorithm but it can show up in the reconstructed
kernel ${\cal K}$ which then appears somewhat jagged. Note also that
the polynomial is intended to produce an interpolation without spurious
oscillations. It is not intended to significantly reduce the errors from
the data, for which a much larger window should be chosen. Suppressing
the errors from the continuum is of secondary importance since this can
better be done by increasing the total length of the time series, and
therefore the number of base functions, which only requires additional
observations. The cost of smoothing with a window with more points is
a degraded limiting resolution unless the sampling is done at
a higher rate. It is certainly impossible to get a higher sampling rate
in the part of a time series that is already obtained. Also there are
practical limitations to the highest sampling rate in observing campaigns,
whereas extending a time series is always possible in principle.

The resolution limit imposed by this scheme is not uniform since it is
proportional to the length in time of the smoothing window. A guide-line
for the smallest value of $\Delta$ is found by equating the FWHM of the
Gaussian target function with the width of the window:
\eqnam\guidel{guidel}
$$
\Delta_{\rm min}\ \sim\ {2N_{\rm s}+1\over2\sqrt{\ln2}}
 \Delta t_{\rm av} > \Delta t_{\rm min}
\eqno\neqn$$
Here $\Delta t_{\rm av}$ is the average time interval between individual
flux measurements. As an example consider a time series of 100 measurements
spread over one year. The limiting value for the resolution width would
be $\Delta \sim 11\ {\rm d}$.

\subsection{Error propagation}

Combining the steps in the previous sections leads to a working inversion
algorithm. Such an algorithm is not much use without an estimate of the
error in the determination of the TF, which are simply propagated
measurement errors. In the original SOLA formulation the error propagation
is relatively simple. The TF is obtained from linear combinations of the
line fluxes and therefore the variance of the TF is simply
\eqnam\LiErr{LiErr}
$$
\sigma^2 (\Psi (\tau_0 ))\ =\ {\bf c}^{\rm T} (\tau_0 )\cdot
{\bf E}_{\rm L} \cdot {\bf c} (\tau_0 )
\eqno\neqn$$
Here ${\bf E}_{\rm L}$ is the error variance-covariance matrix of the
line-flux measurements. Usually the errors in the individual line-flux
measurements are assumed to be uncorrelated which means that the matrix
${\bf E}_{\rm L}$ is diagonal.

In the new formulation of SOLA the coefficients ${\bf c}$ should also
have an error associated with them due to the fact that they are
calculated from the base functions $C$ that are not error-free.
Calculating the error propagation through a matrix inversion is extremely
cumbersome. Furthermore the convolution in Eq. \LFTimSer) implies
that the error estimate should involve the integral of the unknown TF
even if the errors in the individual measurements of the continuum flux
are uncorrelated. Finally, adding error terms in the definition \crocor)
for the cross-correlation matrix of the base functions $C$ means that
the matrix ${\bf E}$ will not be a diagonal matrix. This is not a problem
for the algorithm but the error estimate is likely to be sensitive
to the interpolation scheme.

There is a simpler way to measure the uncertainty due to the measurement
errors in the continuum flux. Once the coefficients ${\bf c}$ have been
obtained they are regarded as known with an arbitrarily high precision.
The coefficients are combined with the line-flux measurements in the
usual way and the error estimate is obtained from \LiErr). The new
element in the analysis is to combine the coefficients with the errors
in the base functions $C$. This will produce an error estimate for
each point of the integration kernel ${\cal K}$, with the propagated
errors obtained from the interpolation scheme and the coefficients.
Calculating this error propagation is relatively simple since it results
entirely from linear combinations with known coefficients of the
least-squares fits of the interpolating polynomials which have associated
error estimates (cf. Eq.~4).

The departure of ${\cal K}$ from the target form ${\cal T}$ is already
used for an estimate of the error due to an imperfect match of the two
(cf. Pijpers \& Thompson, 1994). Instead of a strict upper limit to this
error that would be obtained if the continuum flux was error free, the
upper limit is now `fuzzy'. This arises because there is an error
$\delta {\cal K} (\tau )$ associated with the integration kernel
${\cal K}$. Because the kernel is only used in integrated form the
effect of a variance is reduced. If by analogy to the definition used by
Pijpers \& Thompson (1994) a `systematic error' measure $\chi$ is defined
$$
\chi\ \equiv\ \int\limits_{0}^{\tau_{\rm max}} {\rm d}\tau\
\left[ {\cal K} - {\cal T} \right]^2 ,
\eqno\neqn$$
the upper limit to the systematic error in $\Psi$ is given by~:
$$
E_{\rm upp} = \left( { \Psi_{\rm max} - \Psi_{\rm min} \over 2} \right)
\chi^{1\over 2}
\eqno\neqn$$
where $\Psi_{\rm min}$ and $\Psi_{\rm max}$ refer to the minimum and
maximum value of the TF over the range $[0, \tau_{\rm max}]$.
Only if this error $E_{\rm upp}$ is of the same order of magnitude
as the $\sigma \left( {\cal K} (\tau )\right)$ or smaller than that,
will the errors in the integration kernel significantly affect the error
estimate $E_{\rm upp}$ due to imperfect matching of the calculated
integration kernel ${\cal K}$ to the target function ${\cal T}$.

In this way the errors in the data are fully taken into account. The
errors in the line fluxes are propagated properly into an error
estimate for the TF determination. The errors in the continuum fluxes
are propagated properly into an error estimate of the integration kernel
${\cal K}$. The departure of ${\cal K}$ from ${\cal T}$ can be used to
estimate the effect of this on the estimate of the TF at the appropriate
$\tau_0$. It should be noted that this measure $\chi$ is not available
in the classical MOLA methods such as the original Backus \& Gilbert
method (1967, 1968, 1970). This is due to the fact that the concept of
a target function is unique to the SOLA method.

At this point one could argue that the minimization in Eq. \OldMin)
is not optimal if only the errors in the line flux are included in the
matrix ${\bf E}$. From Eq. \LFTimSer) it can be seen that if there is
an uncorrelated measurement error associated with the continuum
measurement as well as the flux measurement, then this will introduce
an additional variance of the order of $\Phi^2 \sigma^2\left(C(t_i)\right)$.
The integral $\Phi$ of $\Psi$ can be determined independently (cf. sec. 4.5).
The combination of the errors in the continuum and emission-line light
curves can be estimated very roughly by adding the two. For the matrix
${\bf E}$ used in the optimization such a rough estimate is sufficient. The
form that is used here is given by~:
\eqnam\errest{errest}
$$
\eqalign{
E_{ii}\ &=\ \left[ \sigma^2 \left( L(t_i) \right) + \Phi^2 \sigma^2
\left( C(t_i) \right) \right] / E_{\rm n} \cr
E_{ij}\ &=\ 0 \qquad\qquad i\ne j \cr}
\eqno\neqn$$
Because of the interpolation of the continuum light curve the errors
${\delta C}$ are not really uncorrelated. It is therefore not correct
to use these estimates for the final error determination. It is therefore
never used for that but instead the analysis is used that is described in
the first part of this section. However, for the minimization in \OldMin)
it is a sufficiently accurate approximation to ensure that the errors
on the constructed kernel ${\cal K}$ remain sufficiently small to produce
meaningful results.

In Eq. \errest) the factor $E_{\rm n}$ is introduced to produce a
dimensionless quantity. In helioseismology it is customary to make the
trace of the matrix ${\bf E}$ equal to the number of base functions.
The same is done here~:
$$
E_{\rm n}\ =\ {1\over N} \sum\limits_{i =1}^{N} \sigma^2 \left(
L(t_i) \right) + \Phi^2 \sigma^2 \left( C(t_i) \right)
\eqno\neqn$$

\subsection{Simultaneity of line and continuum measurements}

In the previous sections it is assumed that the measurements of the line
flux and of the continuum flux are simultaneous. This is not necessary
for the method to work. As was mentioned before at the start of the time
series the line-flux measurements are of little use so they could be
omitted.

If there are continuum-flux measurements not accompanied by a measurement
of the line flux these can simply be incorporated into the base functions
$C$. This does not add a new base function but it does improve locally
the accuracy of the interpolation.

If there are line-flux measurements not accompanied by a measurement of
the continuum flux these can also be added to the data. In this case a
new base function is added with a value at $\tau = 0$ of this base
function that is interpolated between subsequent and successive
measurements of the continuum flux. It is important to note that adding
such line-flux measurements can serve to improve the $S/N$ ratio in the
TF somewhat,
but that they cannot improve the resolution with which the TF can be
determined since the appropriate time resolution in the continuum flux
is absent.

\fignam{\simTF}{simTF}
\beginfigure{2}
\vskip 4.3cm
\caption{{\bf Figure \simTF.} The TF that was used for the simulations}
\endfigure\nfig

\section{Simulations}

\subsection{The set-up}
\fignam{\simlcv}{simlcv}\nfig
\fignam{\siinhi}{siinhi}\nfig
\fignam{\siinlo}{siinlo}\nfig

To test the SOLA method on typical AGN variability data we chose
a representative continuum light curve and a representative TF to
simulate real data. The representative continuum light curve was
taken to be the NGC\thinspace{}5548 light curve as published by Peterson et al.
(1994). This four-year light curve was interpolated linearly onto
a regular grid and then smoothed with a moving-average filter of
total width 40~d. The weights for the individual points were
taken to be equal. Note that this smoothing has implications for the time
resolution that can then be attained in the determination of the TF.

Previous MEM reconstructions of the TF yielded TFs with little or no
response at zero lag and peaking at 10--20~d lag. We thus decided
to use a TF with similar characteristics. We calculated a TF using
the RAN model from Blandford \& McKee (1982) where we chose an
inner radius $r_0$ of the BLR of 15 light days. Thus our artificial
TF peaks at 30~d lag and has little response at zero lag. The
maximum lag $\tau_{\rm max}$ with a non-zero TF is 60~d. This TF was
normalized to unity in order to compare the integral over the TF of the
reconstructions with unity. The resulting TF is shown in Fig. \simTF{}.
The sharp peak may be unrealistic but does provide us with an
estimate of how well the SOLA method can reconstruct sharp features.
The peak was chosen to lie at 30~d instead of 20~d because the
smoothing of the continuum time series degrades the final attainable
resolution and we decided to `stretch' the simulated TF by 50\% in
order to retain approximately the same resolution relative to
$\tau_{\rm max}$ as in the real data where $\tau_{\rm max}\sim 40\ {\rm d}$.

\beginfigure*{3}
\vskip 18.5cm
\caption{\simlcv}{The eight light curves used for the simulations. The left
column shows the continuum light curve and the right column the
emission-line light curve. The upper row shows the noise-free curves,
the second row 2\% noise has been added to the data, the third row
5\% noise, and the lower row 10\% noise}
\endfigure

We then convolved the artificial continuum light curve with this
TF and obtained an artificial emission-line light curve. The two
light curves were then chopped such that only the second year of
data (cf. Peterson et al. 1994) remained. This reduced data set
was then resampled irregularly onto the same grid as the original
observations were done. In this way two ideal (i.e., noise-free)
light curves were produced.

To examine the effects of data errors, random noise from a Gaussian
distribution with a standard deviation of 2\%, 5\% and 10\% of the
signal was added. The routine RAN1 from the Numerical
Recipes (Press et al. 1992) was used as the random-number generator
and for every continuum light curve the same seed was used, such that
the noise pattern was identical for each continuum curve except for
the magnitude of the noise. The same was done for the emission-line
light curves albeit with a different seed to avoid correlated errors
between the continuum and emission-line time series.
The eight light curves (four continuum and four emission-line
with 0\%, 2\%, 5\% and 10\% noise each) are presented in Fig. \simlcv{}.

\beginfigure*{4}
\vskip 19.0cm
\caption{{\bf Figure \siinhi.} The recovered TFs with a five-day resolution for
the simulated
data. The target TF is shown as a dashed line in each plot}
\endfigure

These light curves were then used to test the SOLA method. Ten simulation
runs were devised according to the scheme in Table 1. In simulations 1--4 the
noise was increased from 0 to 10\% on both light curves. These simulations
reflect real data. In simulations 5--7 only the noise in the emission-line
light curve was increased from 2 to 10\% in order to test the effect of
the errors on the emission-line data only, and in simulations 8--10 only
the noise in the continuum was increased from 2 to 10\% whereas the
emission-line light curve was taken to be error-free. These last
simulations thus measure the effect of continuum-error propagation.

\begintable{1}
\caption{{\bf Table 1.} The simulations}
\halign{#\hfil&\quad#\hfil&\quad#\hfil&\quad#\hfil&\quad#\hfil\cr
\noalign{\hrule\medskip}
&\multispan4 Emission-line~noise\hfil\cr
&\multispan4 \hrulefill\cr
continuum noise& 0\% & 2\% & 5\% & 10\% \cr
\noalign{\medskip\hrule\medskip}
0\% & SIM 1 & SIM 5 & SIM 6 & SIM 7 \cr
2\% & SIM 8 & SIM 2 &       &       \cr
5\% & SIM 9 &       & SIM 3 &       \cr
10\%& SIM 10 &      &       & SIM 4 \cr
\noalign{\medskip\hrule}}
\endtable

\subsection{The results}
\fignam{\sikern}{sikern}\nfig

We have run the simulations twice, once with a high resolution of
$\Delta=5\ {\rm d}$ and once with a resolution of $\Delta=10\ {\rm d}$.
The mean sampling interval of the time series is only 3.4~d and one
might argue for being able to reach this resolution but as explained
in Sect. 3 the introduction
of noise prohibits this. For both simulation runs, the trade-off parameter
$\mu$ between the importance of resolution on the one side and error
suppression on the other side was set to 0.001. The results of the
SOLA inversion on the simulated data are presented in Figs. \siinhi{}
and \siinlo{}.
Figure \siinhi{} shows the 5-day resolution solution whereas Fig. \siinlo{}
shows the 10-day resolution solution. As an example, Fig. \sikern{}
displays the kernels of SIM\thinspace{}2 (i.e., the simulation with 2\%
noise on both the continuum and emission-line light curves) with
$\Delta=10\ {\rm d}$.

We used $N_{\rm s}=2$ and
the expected maximum resolution for the error-free data due to the
irregularity of the time series and the interpolation scheme used to
compensate for that is $\Delta_{\rm min}\sim10\ {\rm d}$ (cf. Eq. 12).
We thus expect the simulation with $\Delta=5\ {\rm d}$ to be undersampled.

The first thing we note from Figs. \siinhi{} and \siinlo{} is that the
error bars on the $\Delta=5$ simulation are systematically larger than
on the $\Delta=10$ simulation. This is an obvious result due to the
fact that less light-curve data are used per TF resolution element
because the kernels $\cal K$ for the $\Delta=5$ simulation are narrower
than for the $\Delta=10$ simulation.

The second thing we note in both ($\Delta=5$ and $\Delta=10$) ideal
simulations (SIM\thinspace{}1) is that the TF is recovered to within the
error bars but shows a slightly displaced peak towards shorter $\tau$.
This is due to the asymmetry of the used TF and the skewness of the kernels
at $\tau\la 20\ {\rm d}$. The skewness of the kernel can be seen from Fig.
\sikern{} where the kernels of SIM\thinspace{}2 with $\Delta=10\ {\rm d}$
are displayed. The kernels that are significantly affected, at $\tau=15\
{\rm d}$ and 20~d, are notably asymmetric: they have more power towards
larger lag than shorter lag. Together with the asymmetric simulated TF used
it is obvious that this would raise the reconstructed TF above the input TF
at these lags.

\beginfigure*{5}
\vskip 19.0cm
\caption{{\bf Figure \siinlo.} The recovered TFs with a ten-day resolution for
the simulated
data. The target TF is shown as a dashed line in each plot}
\endfigure

The $\tau=0$ boundary of the reconstructed TF is slightly raised compared to
the input TF, especially in the $\Delta=10$ case,
because there is no data on the negative side of the boundary
and the finite width of the kernels let more-positive signal from larger
$\tau$ leak into the $\tau=0$ domain. If the peak of the TF would lie
close to a resolution element away from $\tau=0$, this boundary effect
will pull the observed peak of the TF towards $\tau=0$, resulting in an
underestimation of the size of the BLR. The effect of the boundary is made
clear by the kernels in Fig. \sikern{}. Here it is observed that with
$\Delta=10\ {\rm d}$ the kernel at $\tau=5\ {\rm d}$ is still effectively
a kernel at $\tau=0\ {\rm d}$ and also the kernel at $\tau=10\ {\rm d}$
is significantly affected by the boundary. This boundary effect must be
taken into account when examining short lags in reconstructed TFs.

\beginfigure*{6}
\vskip 19.0cm
\caption{{\bf Figure \sikern.} The kernels for simulation 2 with a resolution
$\Delta=10\ {\rm d}$ }
\endfigure

Having established these artificial effects on the ideal data set
(SIM\thinspace{}1) we can increase the noise on both continuum and
emission-line light
curves to 2\%, 5\% and 10\% via simulations 2--4. The reconstructed TFs
show increasing error bars but also substructure that must be entirely
due to noise. In the undersampled $\Delta=5$ simulations we would hardly
recognize the TF in the 5\%-noise case (SIM\thinspace{}3) even though the
reconstruction is consistent with the input TF to within the errors.
SIM\thinspace{}3 is still acceptable in the $\Delta=10$ case though also
here we undersample because the noise degrades the maximum attainable
resolution of 10~d somewhat. The real data on NGC\thinspace{}5548 have
error bars on the 3\% level, close to SIM\thinspace{}2 and significantly
better than SIM\thinspace{}3. For the real data, a resolution of
$\Delta\sim10$~d would be expected with $N_{\rm s}=2$. Even though the
sampling rate of the NGC\thinspace{}5548 light curves is
significantly better than 10~d it is not legitimate to aim to resolve
structure in the TF on scales less than $\sim$10~d! Actual experimenting
with $N_{\rm s}$ and $\Delta$ on the real data suggested $N_{\rm s}=2$
and $\Delta=8$ is the preferred parameter setting for the NGC\thinspace{}5548
data.

\fignam{\sireso}{sireso}\nfig  % figuur volgt iets later

Simulations 5--7 show the effect of increasing the noise on the
emission-line light curve while using the error-free continuum data.
It is clear that this will increase the error bars and substructure
considerably and it merely shows that the resolving power of the
inversion degrades, as expected, and we aim for a too high resolution
when we set $\Delta =10\ {\rm d}$. The error bars in SIM\thinspace{}7
are larger than in SIM\thinspace{}4 because of the way the data errors
are used in the minimization. In the minimization the variance of the
continuum measurements and line measurements are added. The error bars
in the TF are calculated on the basis of the line-flux measurements only
whereas the errors in the continuum flux are used to obtain an error
estimate for the integration kernel. So with the same weighting factor
$\mu$ the trade-off in the minimization shifts towards smaller error
variance in the TF from SIM\thinspace{}7 to SIM\thinspace{}4. The error
variance in the constructed integration kernel ${\cal K}$ is larger in
SIM\thinspace{}4 than in SIM\thinspace{}7 (cf. Fig.~\sireso{})
which compensates for the decrease of $\sigma \left( {\rm TF} \right)$.

The simulations that used error-free emission-line light curves but
increasingly noisy continuum light curves (simulations 8--10) show
the effect of the error propagation `under the integral'. These
errors clearly introduce spurious signals and structure in the TF
on seemingly random time scales larger than the resolution. This
is a problem that one must test for in order to interpret a
noisy inversion correctly. The effect, however, is moderate up to the
2--3\% noise level.

\subsection{The resolution of the TF}

Figure \sireso{} presents all kernels peaking at $\tau=30\ {\rm d}$ for
both the $\Delta=5$ (left column) and $\Delta=10$ simulations (right
column). These are the kernels responsible for picking up the TF signal
at $\tau=30\ {\rm d}$. Obviously, these kernels have finite width and
the full width at half maximum (FWHM) is a measure of the resolution of
the reconstructed TF. Note that the FWHM is at best $2\Delta\sqrt{\ln{2}}$.
Equation \guidel) shows that with the present interpolation scheme,
the maximum resolution attainable is $\Delta_{\rm min}\sim10\ {\rm d}$
for the present data set, even in the absence of noise! The
FWHM of the kernel for the $\Delta=5$ SIM\thinspace{}1 simulation is
$\sim\!30\ {\rm d}$, implying $\Delta_{\rm min}\sim18\ {\rm d}$. This is
due to the fact that we cannot resolve high-frequency structure that is not
present; recall that these simulations are smoothed versions of a real
light curve with a smoothing window of 40 days. For ideal data, the
interpolation can be done better than done at present because $N_{\rm s}$
can be decreased.

The kernels clearly show that increasing the noise on the continuum-flux
data increases the uncertainty in the determination of the kernel, whereas
the emission-line noise has no effect on the determination of the kernels.
When increasing the noise, the kernels for the $\Delta=5$ simulations
become narrower than the theoretical limit of $\Delta_{\rm min}$. The
reason for this is that the noise is then used by the method just as
real rapid variations would be used : to produce a small width of the
kernel. The noise then allows a closer resemblance of the kernel to the
target function. Thus, in the presence of noise, aiming for a resolution
better than $\Delta_{\rm min}$ means risking spurious structure to show
up in the final TF, because a narrow kernel also means fewer (noisy!)
data points are used for the inversion. A spurious datum can thus have
a large impact on the inversion when aiming for too high a resolution.
This is effectively equivalent to undersampling the data. Comparing the
$\Delta=5$ simulations with the $\Delta=10$ simulations clearly shows
this undersampling problem is not severe if we confine ourselves to
$\Delta\sim\Delta_{\rm min}\sim 10\ {\rm d}$.

Note that aiming for a $\Delta$ less than $\Delta_{\rm min}$ does {\it not\/}
increase the error estimate of the TF. The error bars on the TF are
calculated from the emission-line profiles only, whereas $E_{\rm upp}$ is
calculated from the mismatch of the kernel and the target function. in
the case of noisy data, the noise allows the kernel to closely resemble
the target function even in the absence of intrinsic high-frequency
variations (cf. Fig.~7), thereby keeping $E_{\rm upp}$ small. With
noise-free data, it is not possible to construct a kernel that is
narrower than the minimum variability time scale of the intrinsic
variations. Aiming for $\Delta<\Delta_{\rm min}$ thus equals risking
to fit noise instead of signal; this leads to the spurious substructure
in the TF, and this is evidently not a wise thing to do.

\beginfigure*{7}
\vskip 17.0cm
\caption{{\bf Figure \sireso.} The kernels for all simulations for $\tau=30\
{\rm d}$.
The left column displays the 5-day resolution case; the right column the
10-day resolution case. An artificial vertical offset has been given
for displaying purposes}
\endfigure

\subsection{The cross-correlation function}

It is well-known that the position of the peak of the cross-correlation
function (CCF) of the continuum and emission-line light curves is a
measure of the size of the BLR (Gaskell \& Sparke 1986; Gaskell \& Peterson
1987; Edelson \& Krolik 1988; Robinson \& P\'erez 1990). It is not
difficult to work out (e.g., Bracewell 1986; Penston 1991) that the CCF is
the convolution of the TF with the auto-correlation function (ACF) of the
continuum time series. Because the ACF is a symmetric function, asymmetric
properties of the TF, like e.g. the position of the peak, must be
reflected in a similar way in the CCF. Of course, the reverse is also true.
The error analysis described in the previous sections, can thus also be
applied, at least qualitatively, to the CCF.

Table 2 presents the peak values and the peak positions of the CCFs of
the time series used for SIM\thinspace{}1--SIM\thinspace{}10. It is clear
that increasing the noise on either the continuum or emission-line light
curve decreases the correlation coefficient of the peak of the CCF; noise
increases the width of the ACF and the CCF peak becomes less well-defined.
This is confirmed by Table~2. Also, noisy data decrease the correlation
coefficient at small delays (provided the noise on the continuum and
emission-line light curves are not positively correlated),
thereby increasing the position of the peak
of the CCF towards larger delays. This is also confirmed by Table~2.
Thus, noisy data most likely overestimate the size of the BLR when
only the peak of the CCF is considered. Note that theoretical modelling
of BLRs show the peak of the CCF is more sensitive to the inner boundary
of the BLR than to its centre, thereby underestimating the actual size
of the BLR (Robinson \& P\'erez 1990). However, this is different from
the numerical effect presented here.

Because negative lags are unphysical, with finite resolution not only the
TF, but also the CCF will show a boundary effect at $\tau=0$. The
correlation coefficient at $\tau=0$ will thus be increased for TFs
peaking away from zero delay, and this effect will be larger for
worse resolutions. This pulls the peaks of the TF and CCF towards
shorter lags. In the data set examined here this might
be a problem, especially with the $\Delta=10$ resolution, where the
kernels are significantly asymmetric up to $\tau\sim20\ {\rm d}$ due
to the boundary. Also, the asymmetric TF together with the finite
resolution will shift the peak towards smaller lag.
The peak of the TF is known to be at $\tau=30\ {\rm d}$
but both the TF and the CCF peak at $\tau\sim 23\ {\rm d}$ for the
noise-free data due to the low resolution of the experiment.

Noisy and incomplete data thus reduce the resolving power of both the
TF and the CCF such that the peak positions of the TF and CCF are
displaced.
\begintable{2}
\caption{{\bf Table 2.} The cross-correlation results and the integral
$\Phi$ of the TF
for the simulations. The error in the peak position of the CCF is about
2~d}
\halign{#\hfil&\quad#\hfil&\quad#\hfil&\quad#\hfil&
\quad#\hfil\cr
\noalign{\hrule\medskip}
& CCF($\tau_{\rm peak}$) & $\tau_{\rm peak}$
& $\tophat\Phi$ & $\delta\tophat\Phi$ \cr
\noalign{\medskip\hrule\medskip}
SIM 1 & 0.91 & 23 d & 1.000 & 0.1\% \cr
SIM 2 & 0.89 & 23      & 0.998 & 0.3\% \cr
SIM 3 & 0.79 & 25      & 0.996 & 0.7\% \cr
SIM 4 & 0.62 & 37      & 1.002 & 1.4\% \cr
SIM 5 & 0.90 & 23      & 1.000 & 0.3\% \cr
SIM 6 & 0.85 & 29      & 1.000 & 0.7\% \cr
SIM 7 & 0.74 & 34      & 1.000 & 1.4\% \cr
SIM 8 & 0.90 & 23      & 0.997 & 0.1\% \cr
SIM 9 & 0.85 & 30      & 0.993 & 0.1\% \cr
SIM 10& 0.73 & 30      & 0.993 & 0.1\% \cr
\noalign{\medskip\hrule}}
\endtable

\subsection{The integral of the TF}

As Wanders (1994) showed, the delay integral of the TF is a
physically relevant parameter $\Phi=\int{\rm d}\tau\,\Psi(\tau)$ that
measures the total observed line reprocessing in the BLR. $\Phi$ is
not directly reproducible by the MEM; only by first solving for
the TF and then integrating.

{}From Eq.~\OldInv) it can be seen that an estimate $\tophat\Phi$ of
the integral $\Phi$ can be obtained by letting the kernel $\cal K$
be unity over $\tau\in[0,\tau_{\rm max}]$. This can be accomplished
by aiming for an infinitely low resolution, i.e., by setting $\Delta$
equal to a very large number. In this way, SOLA optimizes correctly
for the integral problem. The calculation can be done at any $\tau_0$
with this flat kernel. Similarly to the calculation of
$\tophat\Psi(\tau_0)$, a direct error estimate for $\tophat\Phi$ is
obtained. The results for the ten simulations are presented in Table~2.
Except for the unrealistic case of error-free line flux measurements
the deviation from 1 of the integral is smaller than the error estimate
from SOLA based solely on the line-flux errors. This means that the errors
in the continuum flux do not prevent an accurate measurement of the integral.
It is immediately clear that $\Phi$ can be recovered to a precision
better than $\sim$1\% for realistic data with $\la$5\% noise. For the
real data on NGC\thinspace{}5548, which have characteristics close to
SIM\thinspace{}2, we can therefore expect an accuracy of $\tophat\Phi$ of
$\sim$1\%.

Constant contributions to the fluxes can confuse this result but will not
be discussed here because they do not influence the conclusions of
the present paper. See Wanders (1994) for a discussion of this.

\section{NGC\thinspace{}5548}

Peterson et al. (1994) published four years of monitoring of
NGC\thinspace{}5548
and presented MEM-recovered TFs for each year of monitoring. We have
taken the same data set and recovered the TF of each year
using the SOLA method. We did
not aim for a higher resolution than $\Delta=8$~d since this would
introduce local
structure in the TF that is due to measurement errors in the continuum
and emission-line light curves.
The trade-off parameter $\mu$ was set to 0.001 for all four
inversions, and the maximum $\tau$ for which we calculated the TF
was 50~d. Runs with larger $\tau_{\rm max}$ showed no significant structure
at larger lags. Since we have to `throw away' the first data of the
the emission-line light curve up to $\tau_{\rm max}$, decreasing
$\tau_{\rm max}$ means including more data.

\begintable{3}
\caption{{\bf Table 3.} The integral $\Phi$ of the TF for NGC\thinspace{}5548}
\halign{#\hfil&\quad#\hfil&\quad#\hfil&\quad#\hfil&\quad#\hfil\cr
\noalign{\hrule\medskip}
year& $\Phi_{\rm SOLA}$ & $\delta\Phi_{\rm SOLA}$ & $\Phi_{\rm av}$ \cr
\noalign{\medskip\hrule\medskip}
1989 & 0.863 & 0.004 & 0.876 \cr
1990 & 0.797 & 0.004 & 0.825 \cr
1991 & 0.776 & 0.007 & 0.794 \cr
1992 & 0.720 & 0.006 & 0.738 \cr
\noalign{\medskip\hrule}}
\endtable

\fignam{\truinv}{truinv}
\beginfigure*{8}
\vskip 18.5cm
\caption{{\bf Figure \truinv.} The four transfer functions for the
H$\beta$ emission-line
region of NGC\thinspace{}5548. The TFs are for the monitoring years 1989,
1990, 1991 and 1992. Also shown are the upper limits $E_{\rm upp}$ to the
systematic error made due to the uncertainty in the kernels (cf. Eq. 15)}
\endfigure\nfig

The results for the four individual years are shown in Fig. \truinv{}. These
TFs resemble the TFs derived by the MEM (Peterson et al. 1994): little
response at zero lag, little response at lags larger than about 40~d
and peaking at approx. 12--20~d. The decreasing trend in the
position of the peak of the TF from $\sim\!20\ {\rm d}$ in 1989 to
$\sim\!13\ {\rm d}$ in 1992 as was reported by Peterson et al. is also
reproduced here (cf. Fig. \truinv{}).
The peak of the TF is always more than a resolution element away
from the boundary and
therefore the boundary will not have too much influence on the position
of the peak. However, the simulations have shown that a systematic
underestimation of the position of the peak of $\sim$5--7~d cannot
be excluded. The upper limit $E_{\rm upp}$ to the systematic error
due to the uncertainty in the kernels is significantly smaller
than the statistical uncertainty propagated from the line measurements,
indicating we do not run into serious trouble due to the uncertainties
in the continuum measurements.
Note that the TF is not significantly away from zero at zero lag
in any of the reconstructions, except for the year 1989.
There is no indication that the TF would
be significantly negative anywhere. This does not exclude the TF to become
negative on smaller time scales than the present resolution of
about 10~d, but the positiveness of the TFs suggests the MEM
does not run into serious trouble while using its positivity constraint,
at least not with the resolutions attained with the present data.
Negative values in the TF might arise e.g. due to nonlinear line responses
(Sparke 1993).

Note that the SOLA method is not hampered or aided by a positivity
constraint and negative responses would have been recovered as long as
a convolution of a gaussian with such a negative signature in the TF is
larger than the noise level. Experimenting
with the resolution ($\Delta$) showed that when a too high resolution
was aimed for, artificial structure appeared in the TFs, similar to
the noisy simulations in Sect.~4. These artificial structures can become
significantly negative, but are totally due to noise. Care must be taken
not to overinterpret such structure.

In all, the present SOLA results are consistent with previous MEM
results for the recovery of the shape of the NGC\thinspace{}5548 TF.
This gives us more confidence in the viability of these results and it
presents the SOLA method as an alternative to the MEM.

Table~3 presents the integral of the four TFs of NGC\thinspace{}5548 as derived
by the SOLA method ($\Phi_{\rm SOLA}$) and as published by Wanders (1994)
who derived $\Phi$ via the averaging method $\Phi_{\rm av}=\langle L(t)
\rangle / \langle C(t) \rangle$. The errors on $\Phi_{\rm av}$ are
approximately 2.5\% and as discussed in Sect. 4.5, the errors on
$\Phi_{\rm SOLA}$ are on the order of 1\%.
It can thus be seen that the decreasing trend in $\Phi_{\rm av}$ is confirmed
by $\Phi_{\rm SOLA}$.

A decrease in $\Phi$ is equivalent to a decrease in observed reprocessing
of the emission-line within the BLR (Wanders 1994). We thus confirm the
findings that both the position of the peak of the TF and the total integral
of the TF are time-variable on a time scale of a few years, corresponding
to the dynamical time scale of the BLR clouds. This strongly suggests
evolution of the BLR takes place in NGC\thinspace{}5548 and the system is not
stationary on these time scales.

\section{R\'esum\'e}

This paper is primarily concerned with the development and testing of a
new method for the inversion of time-series of AGN continuum and
line-emission fluxes, to obtain the transfer function of the broad-line
region. The results of the
application of this method to real data of NGC\thinspace{}5548 are also shown.
The various aspects of the method, of the tests and of the applications
are discussed extensively already in sections 3, 4, and 5 respectively
so we will confine ourselves to a few summarizing statements.

The SOLA method constructs an integration kernel from a linear combination
of subsections of the continuum light-curve. The TF is reconstructed
from a linear combination of the emission-line measurements corresponding
to the continuum subsections, using the same linear coefficients. The
value of the TF thus obtained is a weighted average of the TF where the
weighting function is the constructed integration kernel.

The method is not subject to any regularity constraint or positivity
constraint for the TF.

The method allows a very clear determination of the time resolution with
which the transfer function can be determined. Most of the procedure
depends only on the sampling strategy and the standard deviations in
the flux measurements. It is therefore possible to determine the resolution
before reconstructing the TF. This provides a-priori a criterion for
avoiding or discarding spurious sub-structure in the TF.

The method uses the variance in the data to produce the smallest possible
standard deviation in the TF at the time resolution used. Because the
method is linear the calculation of error propagation is particularly
straightforward.

Because only one large-matrix inversion has to be done, the SOLA method
is cost-effective and a typical inversion of the NGC\thinspace{}5548 data
(typically $\sim$80 base functions) takes approximately one hour of CPU
time on a SUN Sparc-2 workstation. This allows for experimenting in parameter
space by varying the control parameters $\mu$ and $\Delta$. In the
original Backus-Gilbert formulation the large number of matrix inversions
prevents parameter-space exploration.

The tests of the method show that for a realistic TF and time sampling
the TF can be recovered to within the formal error estimates.
Because of the finite resolution of the time series some systematic
shifts are found in the position of the peak of the TF. We argue that
this is essentially an effect of resolution. It is therefore
intrinsic to the problem and the sampling. This means that other methods
to determine this peak position such as the CCF and the MEM must also
suffer from these systematic effects.

The application of the method to the data obtained for NGC\thinspace{}5548
confirms
the time evolution found previously of the TF over the four years 1989-1992:
both the position of the peak of the TF and its total integral $\Phi$
are time-variable. This evolution of the broad-line region takes place
on the dynamical time scale of the clouds within the BLR.

The results of the method, the tests, and the application to real data
allow us to make precise recommendations for setting up observing
campaigns, based on the required resolution of the TF and the expected
standard deviation in the flux measurements. The strategy of current
observing campaigns cannot be expected to produce much higher resolution
or better results than shown here and we strongly urge future monitoring
campaigns to use a higher sampling rate and an even better photometric
accuracy. Unfortunately, both will increase the pressure on telescope
time.

The method can easily be adapted to a 2-D inversion for the TF as a
function of time-lag and velocity. Considering that the shape of an
emission-line profile can be determined more accurately than its flux
this may yield a great deal of new information. For instance, the SOLA
method opens the way to the determination of the $Q$-function as introduced
by Perry, van Groningen \& Wanders (1994). They split the full TF into
a steady-state and a deviation term:
\eqnam\Qparam{Qparam}
$$
\Psi(v,\tau)={\Phi(v)\over\Phi}\Psi(\tau)+Q(v,\tau)\eqno\neqn
$$
where $\Phi(v)=\int{\rm d}\tau\,\Psi(v,\tau)$ is the steady-state profile,
i.e., the emission-line profile in response to a constant continuum
flux. $Q(v,\tau)$ is then a function extremely sensitive to variations
in the emission-line profile and from this a great deal about the
structure of the BLR can be learned. Perry et al. present an inversion
problem to directly measure $Q(v,\tau)$ via normalized emission-line
profiles. Because $Q(v,\tau)$ is partly negative, the MEM cannot recover
this function because of its built-in positivity constraint. This is
no problem for the present SOLA method.

Also the steady-state profile $\Phi(v)$ is an important quantity and
can be measured from variable sources by the SOLA method similarly to
the way $\Phi$ is measured as described in Sect. 4.5. Evolution of the
BLR will reflect itself in $\Phi(v)$ also and $\Phi(v)$ will thus be
a powerful way to study evolution while simultaneously being comparatively
easy to calculate by the SOLA method.

The SOLA method is thus not only an alternative to existing inversion
techniques for the derivation of the TF, but can be expected to be a
powerful tool in the study of emission-line profiles as well, which
opens the way to a better understanding of the broad-line region in
active galactic nuclei.

\section*{Acknowledgements}
We thank Katalin and Rackis for the inspiration given during
the development of our ideas resulting in the present paper. We thank
Keith Horne for his detailed comments which helped to improve
the presentation of this paper.

\bye